\documentclass{osa-article}

\journal{osac}

\articletype{Research Article}

\begin{document}

\title{Large mechanical squeezing beyond 3dB of hybrid atom-optomechanical systems in highly unresolved sideband regime}

\author{Jian-Song Zhang \authormark{1} and Ai-Xi Chen\authormark{2,3,*}}

\address{\authormark{1}Department of Applied Physics, East China Jiaotong University,
Nanchang 330013, People's Republic of China\\
\authormark{2}Department of Physics, Zhejiang Sci-Tech University, Hangzhou 310018, People's Republic of China\\
\authormark{3}Institute for Quantum Computing, University of Waterloo, Ontario N2L3G1, Canada}

\email{\authormark{*}aixichen@zstu.edu.cn} 



\begin{abstract}
We propose a scheme for the generation of strong mechanical squeezing beyond 3dB in hybrid atom-optomechanical systems
in the highly unresolved sideband (HURSB) regime where the decay rate of cavity is much larger than the frequency of the mechanical oscillator. The system is formed by two two-level atomic ensembles and an
optomechanical system with cavity driven by two lasers with different amplitudes.
In the HURSB regime, the squeezing of the movable mirror can not be larger than 3dB if no atomic ensemble or only one atomic
ensemble is put into the optomechanical system.
However, if two atomic ensembles are put into the optomechanical system, the strong mechanical squeezing beyond 3dB is achieved
even in the HURSB regime. Our scheme paves the way toward the implementation of strong mechanical squeezing beyond 3dB in
hybrid atom-optomechanical systems in experiments.
\end{abstract}

\section{Introduction}
Quantum squeezing of quantum systems is a characteristic property of macroscopic quantum effects \cite{Agarwal2013}.
It can be used to detect weak forces and realize continuous variable quantum information processing \cite{Braunstein2005}.
Quantum squeezing can be achieved using the parametric interaction of a quantum system \cite{Walls2008}. Unfortunately, quantum
squeezing in this scheme can not be larger than 3dB, i.e., quantum noise of a system could not be reduced
below half of the zero-point level \cite{Milburn1981}. Otherwise, the quantum system becomes unstable.

Up to now, several schemes have been proposed to overcome the 3dB limit by
using continuous weak measurement and feedback \cite{Ruskov2005,Clerk2008,Szorkovszky2011,Szorkovszky2013},
squeezed light \cite{Jahne2009,Huang2010},
quantum-reservoir engineering \cite{Rabl2004,Mari2009,Zhang2009,Gu2013},
strong intrinsic nonlinearity \cite{Asjad2014,Lv2015},
auxiliary cavities and atoms \cite{Wangdongyang20191,Wangdongyang20192},
two driving lasers with different
amplitudes \cite{Kronwald2013,Wollman2015,ZhangRong2019}, and
frequency modulation \cite{Han2019}.

For instance, strong mechanical squeezing beyond 3dB can be realized by injecting a squeezed light into an
optomechanical system since the optical squeezing of the squeezed light can be transferred into
the mechanical resonator due to the interactions between the cavity and mechanical resonator \cite{Jahne2009,Huang2010}.
Also, the authors of Ref. \cite{Lv2015} have shown that the
Duffing nonlinearity can be used to accomplish strong mechanical squeezing (beyond 3dB) in
optomechanical systems when the nonlinear amplitude is large enough.
In Ref. \cite{Kronwald2013}, arbitrarily large steady-state mechanical squeezing can be realized by applying
two driving lasers with different amplitudes to a cavity in an optomechanical system.
In 2015, the squeezing of mechanical mode is realized experimentally by the authors of Ref. \cite{Wollman2015} using the method proposed by Kronwald, Marquardt, and Clerk \cite{Kronwald2013}.
Note that in the unresolved
sideband regime when the decay rate of the cavity $\kappa$ is larger than the frequency of the mechanical oscillator $\omega_m$,
this scheme is no longer valid. Recently, it has been pointed out the beyond 3dB strong mechanical squeezing can be achieved
with the help of frequency modulation \cite{Han2019}.

Quantum squeezing beyond the 3dB limit has been realized in electromechanical systems in experiment \cite{Lei2016}.
In recent years, many efforts have been devoted to the study of optomechanical systems due to
their wide applications \cite{Bowen2015,Aspelmeyer2014,Xu2016,Lv20152,Yin2017,Xiong2017,Woolley2008,Nunnenkamp2009,Vinante2013,Purdy2013,Pontin2014,Genoni2015,Patil2015,Cai2016,Agarwal2016,Chauhan2016,Bennett2018,Hu2018,Zhang2017,Zhang2019}.
In order to realize strong quantum squeezing beyond the 3dB limit in standard optomechanical systems,
the decay rate of the cavity $\kappa$ must be much smaller than the frequency of the mechanical oscillator $\omega_m$
(resolved sideband regime with $\kappa \ll \omega_m$) \cite{Bowen2015,Aspelmeyer2014}.
In experiments, the quality factor of optical cavities should be very high in order to satisfy the resolved sideband criterion.
This limits the mass and size of mechanical resonator
to be squeezed. The 3dB limit has not been overcome in optomechanical systems
since it is difficult to enhance the quality of cavities in optomechanical systems
with floppy mechanical elements \cite{ZhangRong2019} and the resolved sideband regime is
difficult to achieved in experiments. Very recently, the authors of Ref. \cite{ZhangRong2019}
have shown that quantum squeezing beyond the 3dB limit in the unresolved sideband regime ($\kappa \approx 30 \omega_m$)
can be realized in an optomechanical system with two auxiliary cavities and two lasers.
It is worth noting that in order to realize mechanical squeezing the decay rates of two auxiliary cavities
should be smaller than the frequency of the mechanical resonator \cite{ZhangRong2019}.
In particular, the scheme is not valid in the HURSB regime with $\kappa \gg \omega_m$.

All the previous schemes \cite{Asjad2014,Lv2015,Wangdongyang20191,Wangdongyang20192,Kronwald2013,Wollman2015,ZhangRong2019,Han2019}
are not able to realize strong mechanical squeezing beyond the 3dB limit in the HURSB regime with $\kappa \gg \omega_m$. For instance, in Refs.\cite{Kronwald2013,Wollman2015}, large mechanical squeezing can be generated in electromechanical systems in the resolved-sideband regime. However, it is difficult to achieve the resolved-sideband regime for optomechanical systems. Therefor, the authors of Ref.\cite{ZhangRong2019} have suggested to generate strong mechanical squeezing in unresolved-sideband regime for optomechanical systems with the help of two auxiliary cavities. However, the quality factors of two auxiliary cavities must be high enough.  
In the present work, we propose a scheme to achieve strong mechanical squeezing beyond the 3dB limit
in the HURSB regime
with the help of two two-level atomic ensembles and two driving lasers with different amplitudes. It is feasible to couple atoms to photons of a cavity in experiments
\cite{Camerer2011,Purdy2010,Ritsch2013,Jockel2015,Chenxi2015,Turek2013,Mann2018}. Particularly, the linewidth of
atoms is very narrow and the decay rate of atoms could be much smaller than the frequency of
the mechanical oscillator \cite{Chenxi2015}. The cavity is driven by two lasers with different
amplitudes. There is an optimal ratio of the driving strengths of the lasers. For realistic
parameters, we can obtain mechanical squeezing beyond 3dB even in the HURSB regime.

\section{Model and effective Hamiltonian}

\begin{figure}[htbp]
\centering\includegraphics[width=9cm]{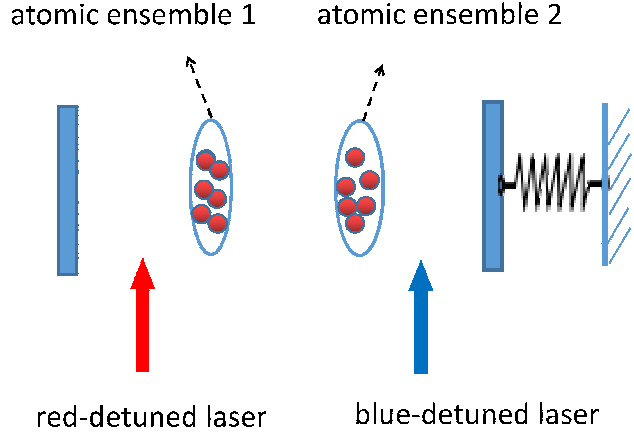}
\caption{The schematic representation of the present model. The cavity is formed by a fixed mirror and a movable mirror.
The movable mirror is perfectly reflecting
while the fixed mirror is partially transmitting. Two ensembles formed by two-level atoms are put into a cavity with frequency $\omega_c$.
The cavity is driven by two lasers with frequencies $\omega_\pm = \omega_c \pm \omega_m$ and amplitudes $\Omega_\pm$.
In experiments, it is difficult to realize high quality factor cavities containing movable mechanical elements to
satisfy the resolved sideband regime ($\kappa \ll \omega_m$).
In the present work, we assume the decay rate of the cavity is much larger than the frequency of the mechanical resonator ($\kappa \gg \omega_m$).} \label{fig1}
\end{figure}

Here, we consider a hybrid atom-optomechanical system consisted of two two-level atomic ensembles
within a single-mode cavity as shown in Fig.1. The cavity is driven by two lasers with different amplitudes $\Omega_\pm$.
The Hamiltonian of the present model is (we set $\hbar = 1$)
\begin{eqnarray}
H &=& \omega_c a^{\dag}a + \omega_m b^{\dag}b + \frac{\omega_1}{2} \sum_{j=1}^{N_1} \sigma_{z,1}^{(j)} + \frac{\omega_2}{2} \sum_{j=1}^{N_2} \sigma_{z,2}^{(j)}\nonumber\\
&& + g a^{\dag}a (b^{\dag} + b) + [g_1 a \sum_{j=1}^{N_1}\sigma_{+,1}^{(j)} + g_2 a \sum_{j=1}^{N_2}\sigma_{+,2}^{(j)}
 \nonumber\\
&& + (\Omega_+ e^{-i\omega_+ t} + \Omega_- e^{-i\omega_- t})a^{\dag} + H.c.],
\end{eqnarray}
where $a^{\dag}$ ($a$) and $b^{\dag}$ ($b$) are the creation (annihilation) operators of the cavity field and mechanical resonator
with frequencies $\omega_c$ and $\omega_m$, respectively.
Here, $\sigma_{z,s}^{(j)} = |e\rangle_s^j\langle e| - |g\rangle_s^j\langle g|$, $\sigma_{+,s}^{(j)} = |e\rangle_s^j\langle g|$,
and $\sigma_{-,s}^{(j)} = |g\rangle_s^j\langle e|$ are the Pauli matrices of atom $j$
in ensemble $s$. The frequencies of ensemble 1 and ensemble 2 are $\omega_1$ and $\omega_2$, respectively.
The numbers of atoms in ensemble 1 and ensemble 2 are $N_1$ and $N_2$, respectively.
The coupling strength between ensemble 1 (ensemble 2) and the single-mode cavity is
$g_1$ ($g_2$). The coupling strength between the cavity and mechanical resonator is denoted by
$g$. Here, $\Omega_\pm$ and $\omega_\pm = \omega_c \pm \omega_m$ are the amplitudes and frequencies of two lasers, respectively.
The decay rates of cavity, mechanical resonator, atomic ensemble 1, and atomic ensemble 2 are denoted by $\kappa$, $\gamma_m$, $\gamma_1$,
and $\gamma_2$, respectively.
The first line of the above equation is the free Hamiltonian of the whole system.
The first term of the second line is the single-photon optomechanical coupling strength.
The second and third terms of the second line are the atom-photon coupling strengths.
The two terms of the last line are the two driving lasers of the cavity.
In the present work, we assume $\Omega_- > \Omega_+$, i.e., the amplitude of the red-detuned laser
is larger than that of the blue-detuned laser.

To simplify the Hamiltonian of the present model, we introduce the operators of
atomic collective excitation modes of atomic ensembles
\begin{eqnarray}
A_1 &=& \frac{1}{N_1}\sum_{j=1}^{N_1}\sigma_{-,1}^{(j)},\\
A_2 &=& \frac{1}{N_2}\sum_{j=1}^{N_2}\sigma_{-,2}^{(j)}.
\end{eqnarray}
In the limit of low-excitation and large number of atoms ($N_1$ and $N_2$),
we have the following commutation relations \cite{Holstein1940,Liu2001,Sun2003,Jin2003,Parkins2006,Genes2008,Ma2013}
\begin{eqnarray}
[A_1, A_1^{\dag}] &\approx& [A_2, A_2^{\dag}] \approx 1, [A_1, A_2] = [A_1, A_2^{\dag}] = 0.
\end{eqnarray}
The Hamiltonian can be expressed in terms of $a$, $b$, $A_1$, and $A_2$ as follows
\begin{eqnarray}
H &=& \omega_c a^{\dag}a + \omega_m b^{\dag}b + \frac{\omega_1}{2}A_1^{\dag} A_1 + \frac{\omega_2}{2} A_2^{\dag} A_2 \nonumber\\
&& + g a^{\dag}a (b^{\dag} + b) + [a^{\dag}(G_{A_1} A_1 + G_{A_2} A_2) + H.c.], \label{H}
\end{eqnarray}
with $G_{A_1} = g_1 \sqrt{N_1}$ and $G_{A_2} = g_2 \sqrt{N_2}$.

The above Hamiltonian can be linearized 
by using the following displacement transformations
$a\rightarrow \alpha + \delta a$, $b\rightarrow \beta + \delta b$,
$A_1\rightarrow \alpha_1 + \delta a_1$, and $A_2\rightarrow \alpha_2 + \delta a_2$ \cite{Bowen2015}.
Using the Hamiltonian of Eq.(\ref{H}) and displacement transformations, we get the following quantum Langevin equations
\begin{eqnarray}
\dot{\alpha} &=& -(i\omega_c + \frac{\kappa}{2})\alpha - ig\alpha(\beta + \beta^*) - i G_{A_1}\alpha_1 -iG_{A_2}\alpha_2 \nonumber\\
&& - i(\Omega_+ e^{-i\omega_+ t} + \Omega_- e^{-i\omega_- t}), \nonumber\\
\dot{\beta} &=& -(i\omega_m + \frac{\gamma_m}{2}) \beta - ig|\alpha|^2, \nonumber\\
\dot{\alpha_1} &=& -(i\omega_1 + \frac{\gamma_1}{2})\alpha_1 - iG_{A_1} \alpha, \nonumber\\
\dot{\alpha_2} &=& -(i\omega_2 + \frac{\gamma_2}{2})\alpha_2 - iG_{A_2} \alpha, \nonumber\\
\delta\dot{a} &=& -i[\omega_c + g(\beta + \beta^*)]\delta a - \frac{\kappa}{2} \delta a - ig\alpha(b^{\dag} + b) \nonumber\\
&& - iG_{A_1}\delta a_1 - iG_{A_2}\delta a_2 + \sqrt{\kappa} a_{in}, \nonumber\\
\delta\dot{b} &=& -(i\omega_m + \frac{\gamma_m}{2})\delta b - ig(\alpha^* \delta a + \alpha \delta a^{\dag}) + \sqrt{\gamma_m} b_{in}, \nonumber\\
\delta\dot{a_1} &=& -(i\omega_1 + \frac{\gamma_1}{2}) \delta a_1 - iG_{A_1}\delta a + \sqrt{\gamma_1} a_{1,in}, \nonumber\\
\delta\dot{a_2} &=& -(i\omega_2 + \frac{\gamma_2}{2}) \delta a_2 - iG_{A_2}\delta a + \sqrt{\gamma_2} a_{2,in},
\end{eqnarray}
where we have neglected the nonlinear terms in the above derivation.

Next, we drop the small frequency shift $g(\beta + \beta^*)$ in the above equations \cite{ZhangRong2019} and get the
linearized Hamiltonian as follows
\begin{eqnarray}
H_1 &=& \omega_c \delta a^{\dag}\delta a + \omega_m \delta b^{\dag} \delta b + \omega_1 \delta b_1^{\dag} \delta b_1
+ \omega_2 \delta b_2^{\dag} \delta b_2 \nonumber\\
&& + g[\alpha^*(t) \delta a + \alpha(t) \delta a^{\dag}](\delta b^{\dag} + \delta b) \nonumber\\
&& + (G_{A_1}\delta a^{\dag} \delta a_1 + G_{A_2}\delta a^{\dag} \delta a_2 + H.c.).
\end{eqnarray}
In addition, we have
\begin{eqnarray}
\alpha(t) & \approx & \alpha_+' e^{-i\omega_+ t} + \alpha_-' e^{-i\omega_- t}, \\
\alpha_\pm' &=& \Omega_\pm/(\pm \omega_m + \frac{\kappa}{2} - \xi_{1,\pm} - \xi_{2,\pm}),\\
\xi_{j, \pm} &=& G_{A_j}^2 /(\omega_{\pm} - \omega_j + i\frac{\gamma_j}{2}).
\end{eqnarray}

In the interaction picture defined by
\begin{eqnarray}
U(t) = \exp{\{-i t [\omega_c(\delta a^{\dag}\delta a + \delta a_1^{\dag}\delta a_1 + \delta a_2^{\dag}\delta a_2) + \omega_m \delta b^{\dag} \delta b] \}},
\end{eqnarray}
we obtain the effective Hamiltonian $H_{eff}$ after some algebra
\begin{eqnarray}
H_{eff} &=& \Delta_1 \delta a_1^{\dag}\delta a_1 + \Delta_2\delta a_2^{\dag} \delta a_2 \nonumber\\
&& +
[G_{A_1} \delta a^{\dag}\delta a_1 + G_{A_2}\delta a^{\dag}\delta a_2 + \delta a^{\dag} (G_+ \delta b^{\dag} + G_- \delta b)\nonumber\\
&& + \delta a^{\dag} (e^{-2i\omega_m t}G_+ \delta b + e^{2i\omega_m t}G_- \delta b^{\dag}) + H.c.], \label{Heff}
\end{eqnarray}
where $\Delta_{1, 2} = \omega_{1,2} - \omega_c$ and $G_{\pm} = g\alpha'_{\pm}$ are the effective
couplings between the cavity and mechanical oscillator. 
Without loss of generality, we assume $G_{\pm}$ are real.

\section{Quantum Langevin equations and solution}
In this section, we derive the quantum Langevin equations of the present model
using the effective Hamiltonian Eq.(\ref{Heff}) of the previous section.

\subsection{Quantum Langevin equations}
The quantum Langevin equations of the system defined by $H_{eff}$ can be written as
\begin{eqnarray}
\delta \dot{a} &=& -\frac{\kappa}{2} \delta a + i f_1(t) \delta b^{\dag} + i f_2(t) \delta b
-i G_{A_1} \delta a_1\nonumber\\
&& - iG_{A_2} \delta a_2 + \sqrt{\kappa} a_{in},\\
\delta \dot{b} &=& -\frac{\gamma_m}{2} \delta b + i f_1(t) \delta a^{\dag} + i f_3(t)
\delta a + \sqrt{\gamma_m} b_{in}, \\
\delta \dot{a_1} &=& -(\frac{\gamma_1}{2} + i\Delta_1)\delta a_1 - iG_{A_1} \delta a + \sqrt{\gamma_1} a_{1,in},\\
\delta \dot{a_2} &=& -(\frac{\gamma_2}{2} + i\Delta_2)\delta a_2 - iG_{A_2} \delta a + \sqrt{\gamma_2} a_{2,in},
\end{eqnarray}
with $f_1(t) = -(G_+ + G_- e^{2i\omega_m t})$, $f_2(t) = -(G_- + G_+ e^{-2i\omega_m t})$,
and $f_3(t) = -(G_- + G_+ e^{2i\omega_m t})$.
Here, $a_{in}$, $b_{in}$, and $a_{j, in}$ are the noise operators of the cavity field,
mechanical resonator, and atomic ensembles, respectively. They obey the following correlation functions
\begin{eqnarray}
\langle a_{in}(t) a^{\dag}_{in}(t')\rangle &=& \delta(t - t'),\nonumber \\
\langle a^{\dag}_{in}(t) a_{in}(t')\rangle &=& 0,\nonumber \\
\langle b_{in}(t) b^{\dag}_{in}(t')\rangle &=& (n_{th} + 1)\delta(t - t'),\nonumber \\
\langle b^{\dag}_{in}(t) b_{in}(t')\rangle &=& n_{th}\delta(t - t'), \nonumber \\
\langle a_{j,in}(t) a^{\dag}_{j,in}(t')\rangle &=& \delta(t - t'),\nonumber\\
\langle a^{\dag}_{j,in}(t) a_{j,in}(t')\rangle &=& 0,
\end{eqnarray}
where $n_{th}$ is the mean thermal excitation number of the mechanical oscillator.

\subsection{Covariance matrix and solution}
Now, we define the following quadrature operators $ X_{O = a,b,a_1,a_2} = (\delta O^{\dag} + \delta O)/\sqrt{2}$
and $ Y_{O = a,b,a_1,a_2} = -i(\delta O^{\dag} - \delta O)/\sqrt{2}$.
The noise quadrature operators are defined as $ X^{in}_{O = a,b,a_1,a_2} = (O_{in}^{\dag} + O_{in})/\sqrt{2}$
and $ Y^{in}_{O = a,b,a_1,a_2} = -i(O_{in}^{\dag} - O_{in})/\sqrt{2}$. From the above quantum Langevin equations, we obtain
\begin{eqnarray}
\dot{\vec{u}} &=& A \vec{u} + \vec{n}, \label{dudt}
\end{eqnarray}
where $\vec{u} = ( X_a, Y_a, X_b, Y_b, X_{a_1}, Y_{a_1}, X_{a_2}, Y_{a_2})^T$ and

\begin{eqnarray}
\vec{n} &=& (\sqrt{\kappa} X_a^{in}, \sqrt{\kappa} Y_a^{in}, \sqrt{\gamma_m} X_b^{in}, \sqrt{\gamma_m} Y_b^{in},
\sqrt{\gamma_1} X_{a_1}^{in}, \sqrt{\gamma_1} Y_{a_1}^{in}, \sqrt{\gamma_2} X_{a_2}^{in}, \sqrt{\gamma_2} Y_{a_2}^{in})^T, \\
A &=&\left(
\begin{array}{cccccccc}
-\frac{\kappa}{2} & 0 & -\Im{(f_{12}^+)} & \Re{(f_{12}^-)} & 0& G_{A_1} & 0 & G_{A_2}\\
0 & -\frac{\kappa}{2} & \Re{(f_{12}^+)}  & \Im{(f_{12}^-)} & -G_{A_1}& 0 & -G_{A_2} & 0\\
-\Im{(f_{13}^+)} & \Re{(f_{13}^-)} & -\frac{\gamma_m}{2}  & 0 & 0& 0 & 0 & 0\\
\Re{(f_{13}^+)} & \Im{(f_{13}^-)} & 0 & -\frac{\gamma_m}{2} & 0& 0 & 0 & 0\\
0 & G_{A_1} & 0 & 0 & -\frac{\gamma_1}{2}& \Delta_1 & 0 & 0\\
-G_{A_1} & 0 & 0 & 0 & -\Delta_1& -\frac{\gamma_1}{2} & 0 & 0\\
0 & G_{A_2} & 0 & 0& 0 & 0 & -\frac{\gamma_2}{2}& \Delta_2 \\
-G_{A_2} & 0 & 0 & 0 & 0 & 0 & -\Delta_2& -\frac{\gamma_2}{2}
\end{array}  \right), \label{A} \nonumber\\
\end{eqnarray}
where $\Re(f)$ and $\Im(f)$ are the real and imaginary parts of a complex number $f$, respectively, and $f_{jk}^{\pm} = f_j(t)\pm f_k(t)$.

The dynamics of the present system can be completely described by a $8\times8$ covariance matrix
$V$ with $V_{jk} = \langle u_j u_k + u_k u_j\rangle/2$. Using the definitions of $V$, $\vec{u}$,
and Eq.(\ref{dudt}), we get the evolution of the covariance matrix $V$ as
\begin{eqnarray}
\dot{V} = A V + V A^T + D, \label{dV}
\end{eqnarray}
with $D$ being the noise correlation defined by $D = diag[\frac{\kappa}{2}, \frac{\kappa}{2}, \frac{\gamma_m}{2}(2 n_{th} + 1), \frac{\gamma_m}{2}(2 n_{th} + 1), \frac{\gamma_1}{2}, \frac{\gamma_1}{2}, \frac{\gamma_2}{2}, \frac{\gamma_2}{2}]$.

\section{Strong mechanical squeezing in the HURSB regime}

\begin{figure}[htbp]
\centering\includegraphics[width=9cm]{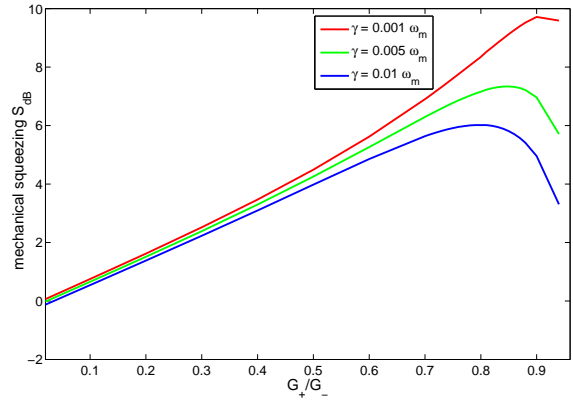}
\caption{The mechanical squeezing $S_{dB}$ are plotted as functions of the ratio $G_+/G_-$ for
$\gamma = 0.001\omega_m$ (red line), $\gamma = 0.005\omega_m$ (green line), and $\gamma = 0.01\omega_m$ (blue line) with
$\kappa = 1000\omega_m$, $\gamma_m = 10^{-5}\omega_m$, $G_{A_1} = G_{A_2} = 10\omega_m$, $G_- = \omega_m$, $n_{th} = 0$,
$\Delta_1 = 2\omega_m$, and $\Delta_2 = -2\omega_m$. The decay rate of cavity is much
larger than the frequency of the mechanical oscillator ($\kappa \gg \omega_m$).
The mechanical squeezing can be larger than 3dB even in the
HURSB regime in the present model as one can see clearly from this figure.} \label{fig2}
\end{figure}

\begin{figure}[htbp]
\centering\includegraphics[width=9cm]{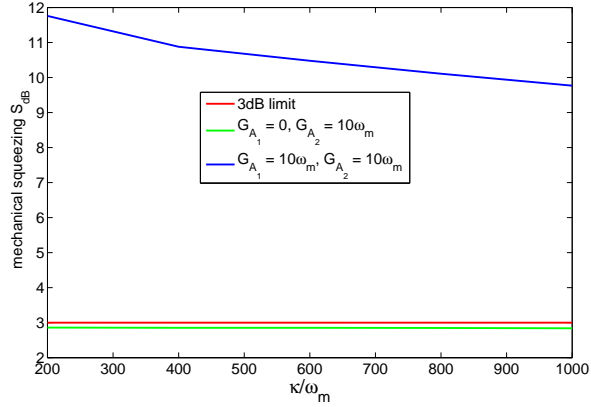}
\caption{The mechanical squeezing $S_{dB}$ are plotted as functions of the ratio of the decay rate of cavity and
frequency of mechanical oscillator $\kappa/\omega_m$ for
$G_{A_1} = 0, G_{A_2} = 10\omega_m$ (green line) and $G_{A_1} = 10\omega_m, G_{A_2} = 10\omega_m$ (blue line) with $G_- = \omega_m$,
$\gamma = 0.001\omega_m$, $\gamma_m = 10^{-5}\omega_m$, $n_{th} = 0$, $\Delta_1 = 2\omega_m$, and $\Delta_2 = -2\omega_m$.
The red line is the 3dB limit.
All the points in the figure have been optimized over $G_+/G_-$.
If no atomic ensemble or only one atomic ensemble is put into the cavity,
the 3dB limit can not be overcome. However, if two atomic ensembles are put into
the cavity, the beyond 3dB mechanical squeezing can be realized even in the HURSB regime.} \label{fig3}
\end{figure}

The mechanical squeezing is defined as (in units of dB) \cite{ZhangRong2019}
\begin{eqnarray}
S_{dB} &=& -10\log_{10}(\langle \Delta X_b^2 \rangle/ \langle \Delta X^2 \rangle_{ZPF}) \nonumber\\
 &=& -10\log_{10}(2 \langle \Delta X_b^2 \rangle),
\end{eqnarray}
where $\langle \Delta X^2 \rangle_{ZPF} = 0.5$ is the zero-point fluctuations.
The mechanical squeezing can be calculated from Eq.(\ref{dV}) of the previous section.

In Fig. 2, we plot the mechanical squeezing (in units of dB) as functions of $G_+/G_-$
for different values of decay rate $\gamma$ (we set $\gamma_1 = \gamma_2 = \gamma$) with $\Delta_1 = 2\omega_m$ and $\Delta_2 = -2\omega_m$.
Here, we assume the system is in the HURSB regime with $\kappa = 1000\omega_m$.
Clearly, the mechanical squeezing of the present work can overcome the 3dB limit
even in the HURSB regime. The mechanical squeezing $S_{dB}$ first increases with
the increase of the ratio $\frac{G_+}{G_-}$ and then decreases with ratio $\frac{G_+}{G_-}$.
There is an optimal ratio $\frac{G_+}{G_-}|_{opt}$. If the blue-detuned laser is not applied
($G_+ = 0$), the 3dB limit can not be surpassed since $S_{dB} < 3$ as one can clearly
see from this figure. Comparing the lines of the figure, we find the optimal ratio $\frac{G_+}{G_-}|_{opt}$ increases with the
decrease of the decay rates of the atomic ensembles.

The influence of the blue-detuned laser can be explained as follows.
In the rotating wave approximation, the direct interactions between the optical cavity and mechanical resonator is represented by the term $\delta a^{\dag}(G_+\delta b^{\dag} + G_-\delta b) + H.c.$ in Eq.(12). Using the standard squeezing transformation \cite{Lv2015}, this term can be rewritten as
$G_{eff}(\delta a^{\dag}\delta B + H.c.)$ with $G_{eff} = \sqrt{G_-^2 - G_+^2}$, $\delta B = \cosh{r}\delta b + \sinh{r}\delta b^{\dag}$ being the Bogoliubov mode, and $r=\ln[(G_- + G_+)/(G_- - G_+)]/2$ being the squeezing parameter. If the blue-detuned laser is not applied ($G_+ = 0$),
the squeezing parameter $r$ is zero and there is no mechanical squeezing. 
In fact, the mechanical squeezing is determined by two competing effects. 
One is the squeezing parameter $r$. The other is the effective direct coupling between 
the optical cavity and mechanical resonator denoted by $G_{eff}$. 
The squeezing parameter $r$ increases with the increase of the ratio $G_{+}/G_{-}$.
However, $G_{eff}$ decreases with the increase of the ratio $G_{+}/G_{-}$.
Consequently, the maximal mechanical squeezing is a tradeoff between these two competing effects. 

In Fig.3, we plot the mechanical squeezing $S_{dB}$
as functions of $\kappa/\omega_m$ for different coupling strengths $G_{A_1}$ and $G_{A_2}$ with
$\gamma = 0.001\omega_m$, $\Delta_1 = 2\omega_m$, and $\Delta_2 = -2\omega_m$.
The 3dB limit is plotted as red line.
From this figure, one can find that if only one atomic ensemble is put into the cavity ($G_{A_1} = 0$, green line),
then the 3dB limit can not been overcome since $S_{dB} < 3$.
In fact, we find that if no atomic ensemble is put into the cavity, the mechanical squeezing
$S_{dB}$ is less than $-30$ in the case of $\kappa \gg \omega_m$ (not shown in this figure).
This implies that the mechanical squeezing beyond 3dB can not
been achieved in the HURSB regime without the atomic ensemble.
The situation is totally different when the two atomic ensembles are put into the cavity as
one can clearly see from the blue line of this figure. In this case, the mechanical squeezing
$S_{dB}$ can be larger than the 3dB limit even in the case of $\kappa = 1000\omega_m$.
The mechanical squeezing $S_{dB}$ decreases with the increase of the ratio $\kappa/\omega_m$.
As one can see from Fig. 2, we can decrease the decay rates of atomic ensembles
$\gamma_1$ and $\gamma_2$ to realize mechanical squeezing beyond the 3dB limit in the HURSB regime.

The two atomic ensembles play an important role in the present scheme. The reason is as 
follows. In Ref. \cite{ZhangRong2019}, two auxiliary high-Q cavities are introduced in order to 
modulate the optical density of states in the cavity. As a result, the the damaging effects of
the counter-rotating terms can be suppressed \cite{ZhangRong2019}. 
In order to overcome the 3dB limit, the decay rates the two auxiliary
cavities must be smaller than the frequency of the mechanical oscillator.
In the present scheme, the two atomic ensembles can also adjust the optical density of states
in the cavity similar to \cite{ZhangRong2019}. In fact, the two
atomic ensembles and the cavity can be considered as an engineered reservoir for 
the mechanical oscillator \cite{Kronwald2013,ZhangRong2019}. 
In particular, there are two main advantages of the present scheme.
First, it is feasible to couple atoms to photons of a cavity field in experiments \cite{Camerer2011,Purdy2010,Ritsch2013,Jockel2015,Chenxi2015,Turek2013,Mann2018}. 
Second, the key requirement of the present scheme is that
the decay rate of atoms must be much smaller than the frequency of
the mechanical oscillator which can be satisfied \cite{Chenxi2015}.  
Thus, the scheme proposed here is feasible in experiments.

\section{Conclusion}
In the present work, we have proposed a scheme to realize strong mechanical squeezing beyond the
3dB limit in the HURSB regime in hybrid atom-optomechanical systems. Two two-level atomic ensembles
were put into the cavity which was driven by two lasers.
The amplitudes of two lasers were assumed to be unequal.
In the limit of low-excitation and large number of atoms, the atomic ensembles can be expressed in terms of bosonic operators. First, we derived an effective Hamiltonian of the present model in the interaction picture.
In the resolved sideband case with $\kappa \ll \omega_m$, the counter-rotating terms of the effective Hamiltonian can be neglected.
However, in the HURSB regime with $\kappa \gg \omega_m$, the influence of the counter-rotating terms
can not been ignored. The dynamics of the present system can be
described by a covariance matrix $V$.
Then, we solved the equation of motion numerically and plotted the mechanical squeezing as functions of the ratio $G_+/G_-$ or $\kappa/\omega_m$.
We found that the 3dB limit of the mechanical squeezing can be overcome even in the HURSB regime. The mechanical squeezing $S_{dB}$ first increases with
the ratio $\frac{G_+}{G_-}$ and then decreases with the ratio $\frac{G_+}{G_-}$.
In particular, the 3dB limit can not be surpassed when the blue-detuned laser is not applied.
In addition, if no atomic ensemble or only one atomic
ensemble is put into the optomechanical system, the squeezing of the movable mirror can not be larger than 3dB.
However, if we put two atomic ensembles into the cavity, the mechanical squeezing beyond 3dB is achieved in the HURSB regime.
Our scheme paves the way toward the realization of large mechanical squeezing beyond the 3dB limit in hybrid atom-optomechanical systems in the HURSB regime.

\section*{Funding}
This research was supported by Zhejiang Provincial Natural Science Foundation of China under Grant No. LZ20A040002; National Natural Science Foundation of China (11047115, 11365009 and 11065007); 
Scientific Research Foundation of Jiangxi (20122BAB212008 and 20151BAB202020.)

\section*{Disclosures}
The authors declare no conflicts of interest.

\bibliography{ref}

\end{document}